\documentclass[aps,pra,reprint]{revtex4-2}
\usepackage{tikz}

\usepackage{color}
\usepackage{graphicx}
\usepackage{epsfig}
\usepackage{epstopdf}
\usepackage{subfigure,color}
\usepackage{dcolumn}
\usepackage{bm}
\usepackage{lineno}
\usepackage{amsmath, amssymb}
\usepackage{accents}
\usepackage{lipsum}
\usepackage{multirow}
\usepackage[usestackEOL]{stackengine}
\usepackage{tabularx}
\usepackage{makecell}
\usepackage{xcolor}
\usepackage{dcolumn}
\usepackage{bm}
\usepackage{booktabs}
\usepackage{comment}
\usepackage{enumerate}
\usepackage{hyperref}
\usepackage{float}
\UseRawInputEncoding

\def\.{\cdot}

\def\##1{{\bf #1\mit}}
\def\_#1{{\bf #1\mit}}
\def\-#1{{\bf #1\mit}}
\def\=#1{\overline{\overline #1}}
\def\Re{{\rm{Re}}}
\def\Im{{\rm{Im}}}
\def\subr{\text{r}}
\def\j{\mathrm j}

\begin{document}
\newcommand{\red}[1]{\textcolor{red}{#1}}

\title{Electromagnetic Effects in Anti-Hermitian Media with Gain and Loss}
\author{L.~Freter$^{1,2}$} 
\author{M.~S.~Mirmoosa$^{3}$}
\author{A.~Sihvola$^1$} 
\author{C.~R.~Simovski$^1$}
\author{S.~A.~Tretyakov$^1$}
	
\affiliation{{$^1$}Department of Electronics and Nanoengineering, Aalto University, P.O.~Box 15500, FI-00076 Aalto, Finland\\{$^2$}Institute of Theoretical Solid State Physics, Karlsruhe Institute of Technology (KIT), 76131 Karlsruhe, Germany\\{$^3$}Department of Physics and Mathematics, University of Eastern Finland, P.O.~Box 111, FI-80101 Joensuu, Finland}


\begin{abstract}
Incorporating both gain and loss into electromagnetic systems provides possibilities to engineer effects in unprecedented ways. Concerning electromagnetic effects in isotropic media that have concurrently electric and magnetic responses, there is in fact a degree of freedom to distribute the gain and loss in different effective material parameters. In this paper, we analytically scrutinize wave interactions with those media, and, most importantly, we contemplate the extreme scenario where such media are anti-Hermitian. Considering various conditions for excitation, polarization, and geometry, we uncover important effects and functionalities such as lasing into both surface waves and propagating waves, conversion of evanescent source fields to transmitted propagating waves, full absorption, and enhancing backward to forward scattering ratio. We hope that these findings explicitly show the potential of anti-Hermiticity to be used in optical physics as well as microwave engineering for creating and using unconventional wave phenomena.
\end{abstract}

\maketitle 


\section{Introduction} 
Non-Hermitian systems have recently engrossed considerable attention in quantum and classical physics~\cite{ashida2020non} including optics~\cite{El2018nonHERM}. Designing non-Hermitian optical systems with balanced distribution of gain and loss allows to have the salient feature of parity-time reversal symmetry~\cite{feng2017NHBOPTSN,Longhi2018PTRS}, which gives rise to a multitude of seductive phenomena and applications such as realizing laser-absorber devices~\cite{Wong2016lasingNL,longhi2010ptLAL, Chong2011_PTLaserAbsorber}, single-mode lasing~\cite{Hodaei2014SMLE,feng2014SMLPTSB}, generation of orbital angular momentum (OAM) lasing~\cite{miao2016OAML}, unidirectional invisibility~\cite{UnidirectionalINPTS, Feng2012_unidirectional, Regensburger2012_unidirectional}, sensing~\cite{hodaei2017ESEPPTSB,Sensing2017EPS, Zhong2019_sensing}, and so forth. Concerning isotropic optical media, parity-time reversal symmetry is achieved by properly engineering the effective permittivity (or dielectric constant) in space such that the real and imaginary parts of this effective parameter are even and odd functions with respect to the position vector, respectively~\cite{ruter2010PTSOP}. However, in contrast to the distribution of the optical gain and loss in space, one alternative way is to distribute the optical gain and loss in different effective material parameters. For linear isotropic media, this is feasible in media possessing simultaneously electric and magnetic responses such as magnetodielectric materials.
If magnetic polarization processes exhibit losses, the electric response can bring equivalent gain, or vice versa. From this point of view, complex conjugate media that have been studied lately~\cite{Dragoman, Xu:17_CCM, Cui2020:RalizationCCM, Basiri2015:CCM} are an example of such media. An interesting question is how light 
interacts with matter in the extreme case of vanishing real parts of permittivity and permeability in such media~\cite{PhysRevA.96.043838_CPA_laser,Bai2016:CCM_absorber,Xu2021:CCM,mirmoosaTellegen}. In this exotic scenario, since the Hermitian transpose of the permittivity and permeability retrieves the permittivity and permeability with a minus sign, the medium is called anti-Hermitian~\cite{mirmoosaTellegen}. In fact, this anti-Hermitian medium is the generalized duality transformation of Tellegen nihility in which the Tellegen parameter is purely imaginary~\cite{mirmoosaTellegen}.

In this paper, we thoroughly investigate electromagnetic wave interactions with anti-Hermitian media and objects. We examine various electromagnetic effects
by studying interfaces, slabs with a back mirror, and spherical inclusions, and, also, by taking different polarizations and excitations into account. We demonstrate that as a result of anti-Hermiticity with nonzero gain and loss, the fundamental properties of  evanescent and uniform plane waves dramatically change, which engenders interesting wave phenomena including a lasing effect of creating surface waves (as well as propagating plane waves), full polarization conversion, perfect matching and absorption, polarization selectivity, and large backward to forward scattering ratio for small inclusions.

The paper is organized as follows: Section~\ref{sec:cartesCG} concentrates on the problems mainly associated with planar geometries (interfaces, slabs, and metasurfaces), and Section~\ref{sec:GSpheG} considers wave interactions with anti-Hermitian spherical particles. Section~\ref{sec:conslW} concludes the paper.


\begin{figure*}[t!]
\centering
\includegraphics[width=1\linewidth]{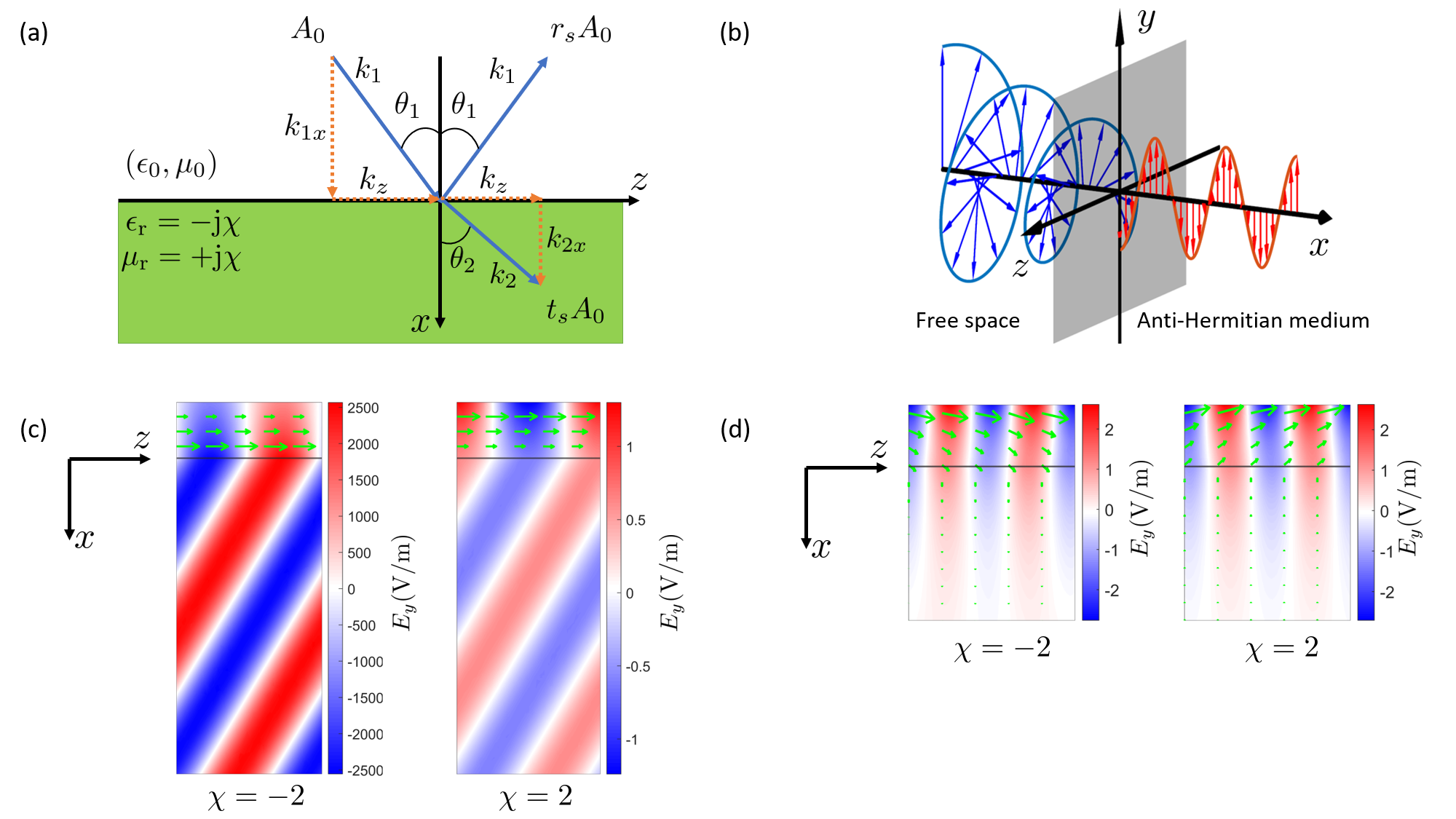}
\caption{(a) Interface between free space and an anti-Hermitian half-space. An incident plane wave with amplitude $A_0$ is partially reflected and transmitted with reflection and transmission factor $r_s$ and $t_s$, where $s=\text{TE,TM}$. (b) Perfect conversion of an elliptically polarized evanescent wave to a linearly polarized propagating wave at an anti-Hermitian interface. (c) Simulation of an interface between free space and anti-Hermitian half-space for an evanescent-wave excitation and a propagating transmitted wave with the material parameter $\chi = \pm2$ and $k_z$ given by Eq.~\eqref{eq:Interface_kz}. The green arrows show the time-averaged Poynting vector. (d) Simulation of an interface between free space and anti-Hermitian half-space for an evanescent-wave excitation and an evanescent transmitted wave for the material parameter $\chi = \pm2$ and $k_z=2.5k_0$. The green arrows show the time-averaged Poynting vector.}
\label{fig:interface}
\end{figure*} 

\section{{Scattering from Planar Geometries}}
\label{sec:cartesCG}
\subsection{Anti-Hermitian interfaces}
We start by considering a planar interface between free space ($x<0$) and an anti-Hermitian half-space ($x>0$) characterized by the relative material parameters
\begin{equation}
    \epsilon_\subr = -\j\chi,\quad \mu_\subr = \j\chi,
    \label{eq:eps_mu}
\end{equation}
in which $\chi$ is a real-valued quantity and $\j$ is the imaginary unit, see Fig.~\ref{fig:interface}(a). In the electrical engineering convention with a time dependence $\mathrm{e}^{\j\omega t}$, a positive value of $\chi$ results in loss in the electric response and gain in the magnetic response, and vice versa. We are interested in the amplitude reflection and transmission factors for a plane wave incident from free space on the interface. By using a plane wave {\it Ansatz} for the incident, reflected, and transmitted fields, and due to the continuity of the tangential components of the fields at the interface $x=0$, we achieve Fresnel's formulae for TE and TM polarized incident waves
\begin{equation}
r_{\rm{TE,TM}}=\frac{\pm\mathrm{j}\chi k_{1x} - k_{2x}}{\pm\mathrm{j}\chi k_{1x}+ k_{2x}},\quad
t_{\rm{TE,TM}}=1+r_{\rm{TE,TM}}.
\label{eq:Interface_r_t}
\end{equation}
These reflection coefficients are defined as ratios of the tangential components of electric fields for TE polarization and of magnetic fields for TM polarization. Here, $k_{1x}=\sqrt{k_0^2-k_z^2}$ and $k_{2x}=\sqrt{n^2k_0^2-k_z^2}$ are the normal components of the wave vectors in the free space and anti-Hermitian half-spaces, respectively, $k_z$ is the transverse wave vector component, $k_0=\omega\sqrt{\epsilon_0\mu_0}$ represents the free-space wavenumber and $n=|\chi|$ is the refractive index of the anti-Hermitian medium. The upper sign in Eq.~\eqref{eq:Interface_r_t} corresponds to  TE polarization, and the lower sign stands for TM polarization. It is evident that in order to switch between the two polarizations, one should exchange $\chi$ by $-\chi$, i.e., $r_{\rm{TE}}(-\chi) = r_{\rm{TM}}(\chi)$. The physical meaning is that TE  and TM  polarized fields behave in exactly the same way if gain and loss in the electric and magnetic responses is flipped. Note that this only holds because in the first half-space the relative permeability and relative permittivity are equal.
Due to the translational invariance of the geometry along the $z$-axis, the transverse component of the wave vector $k_z$ is conserved, and thus identical in both media.

In our study, we always assume $k_z$ to be real, meaning that the incident, reflected, and transmitted waves propagate in the $z$-direction. Depending on the value of $k_z$, the normal components of the wave vectors $k_{1x}$ and $k_{2x}$ are either real, corresponding to a propagating wave, or imaginary, corresponding to an evanescent wave along the $x$-direction. Accordingly, this defines  four different cases: An incident propagating or evanescent wave in free space can be transformed into a propagating or evanescent wave in the 
anti-Hermitian medium. 

Based on Eq.~\eqref{eq:Interface_r_t}, it is clear that if both $k_{1x}$ and $k_{2x}$ are either real or imaginary, the magnitude of the reflection factor $r_{\rm{TE,TM}}$ is unity for both polarizations. An interesting consequence is the fact that in the case of propagating incident and transmitted waves, although the interface shows full reflection, there is propagation without attenuation into the anti-Hermitian medium. Thus, the propagating wave in the anti-Hermitian medium should not carry power, and the time-averaged Poynting vector needs to be zero. For deeper understanding, we write the TE polarized fields as 
$\mathbf{E}=E_{y}\mathrm{e}^{-\mathrm{j}(k_{2x}x+k_zz)}\mathbf{\hat{a}}_y$ and 
$\mathbf{H}=(E_{y}/(\omega\mu_0\mu_\subr))\mathrm{e}^{-\mathrm{j}(k_{2x}x+k_zz)}(-k_z\mathbf{\hat{a}}_x + k_{2x}\mathbf{\hat{a}}_z)$ 
and calculate the time-averaged Poynting vector which is ultimately given by
\begin{equation}
\begin{split}
\langle\mathbf{S}\rangle=\frac{1}{2}\Re[\mathbf E\times\mathbf H^\ast]=\frac{|E_{y}|^2}{2\omega\mu_0\chi}\Im[k_{2x}]\mathrm{e}^{2\Im[k_{2x}]x}\mathbf{\hat{a}}_x,
\end{split}
\label{eq:S_antihermitian_TE}
\end{equation}
where $\Re[...]$ and $\Im[...]$ are the real and imaginary parts of the expression inside brackets. Equation~\eqref{eq:S_antihermitian_TE} explicitly confirms that  the averaged power density in the anti-Hermitian medium indeed vanishes if $k_{2x}$ is real-valued. Interestingly, according to Eq.~\eqref{eq:S_antihermitian_TE}, we uncover that in contrast to propagating plane waves, evanescent waves do carry time-averaged power 
since the corresponding normal component of the wave vector is imaginary for those waves. From a technical point of view, it means that, counterintuitively, the wave impedance is purely real for evanescent waves in an anti-Hermitian medium. This finding will be important later, particularly for explaining full absorption and lasing. It is worth noting that for TM polarization, in Eq.~\eqref{eq:S_antihermitian_TE} we need to exchange $(E_y,\mu_0,\chi)\rightarrow(H_y,\varepsilon_0,-\chi)$, meaning that the qualitative result stays the same.  

If an incident evanescent wave in free space is transformed into a propagating plane wave in the anti-Hermitian medium or vice versa, i.e., $k_{1x}$ and $k_{2x}$ are \emph{not} both real or both imaginary, the reflection factor Eq.~\eqref{eq:Interface_r_t} is purely real, and its magnitude is always larger or smaller than unity, depending both on the sign of $\chi$ and the polarization of the incident wave. In the following, we will investigate each of the four possible cases. We will mainly focus on evanescent-wave excitation, because to the best of our knowledge, these cases have not been discussed before. 


Let us first assume the case of an evanescent-wave excitation in free space that is transformed into a propagating transmitted wave in the anti-Hermitian medium, meaning that  $k_0<k_z<|\chi|k_0$. We will denote this case as \mbox{e $\rightarrow$ p}. It follows that $k_{2x}$ is real and positive, and $k_{1x}=-\mathrm{j}|k_{1x}| = -\mathrm{j}\sqrt{k_z^2-k_0^2}$ is purely imaginary, where the negative sign ensures decaying fields for $x\rightarrow \infty$. Using Eq.~\eqref{eq:Interface_r_t} and  considering only TE polarization (recall that TM polarization solutions we obtain by simply exchanging $\chi$ with $-\chi$), we find
\begin{equation}
    \label{eq:interface_r_ep}
    r_\mathrm{TE}^{\mathrm{e}\rightarrow\mathrm{p}} = \frac{\chi |k_{1x}|-k_{2x}}{\chi |k_{1x}| + k_{2x}}.
\end{equation}
We immediately see that the reflection factor Eq.~\eqref{eq:interface_r_ep} is purely real, and its magnitude is always larger (smaller) than one for negative (positive) $\chi$. For $k_{2x} = \chi |k_{1x}|$, or equivalently for 
\begin{equation}
    k_z = k_0\sqrt{\frac{2\chi^2}{\chi^2+1}} \label{eq:Interface_kz},
\end{equation}
the reflection factor is zero if $\chi$ is positive and has a pole if $\chi$ is negative. 

In the case of zero reflection, the interface shows the remarkable property of perfectly matching an evanescent, exponentially decaying wave to a propagating wave with a constant amplitude. This is not possible using conventional dielectric media, where the reflection coefficient is always of magnitude one in such a case. Moreover, considering a TM polarized wave, the electric field is in general elliptically polarized in free space but linearly polarized in the anti-Hermitian medium, meaning that this interface allows a lossless transformation between elliptical and linear polarizations. This case is illustrated in Fig.~\ref{fig:interface}(b).

On the other hand, in the case of infinite reflection, the interface acts as a laser creating evanescent  waves that exponentially decay into  vacuum. In other words, an evanescent incident wave matching the condition Eq.~\eqref{eq:Interface_kz} results in an (ideally infinitely) enhanced surface wave that carries power only along the $z$-axis. It appears that this is a unique structure that lases directly into a surface wave.  Importantly, the semi-infinite space filled by an anti-Hermitian medium can be terminated at any depth by a matched boundary that is purely reactive, because the wave impedance in this medium is purely imaginary. Consequently, this surface-wave laser can be realized as a thin (as thin as desired) layer of an anti-Hermitian medium on a high-impedance surface. For optical applications, the impedance boundary can be realized as a simple low-loss dielectric slab on a mirror surface. For microwave applications, there are more compact realizations, such as  mushroom layers, for example.

We confirm these results with a numerical simulation performed with the software tool COMSOL Multiphysics, see Fig.~\ref{fig:interface}(c). The anti-Hermitian parameter is set to $\chi=\pm2$, and the interface is illuminated by a TE polarized plane wave with the $z$-component of the wavevector given in Eq.~\eqref{eq:Interface_kz}. For negative $\chi$, the interface acts as a surface laser, manifesting its functionality in strong evanescent fields close to the interface in free space, and a propagating wave with theoretically infinite amplitude in the medium. The time-averaged Poynting vector, shown by green arrows, points only along the $z$-axis in free space and reaches its highest value at the boundary. Inside the medium, the time-averaged Poynting vector is zero. For positive $\chi$, the reflection factor is zero, which is seen in lower values of the electric field and the Poynting vector decreasing in free space along the $x$-direction.


Let us continue with the case of an evanescent-wave excitation that is also evanescent in the anti-Hermitian medium, i.e., $k_z > k_0$ and $k_z>|\chi|k_0$, which we denote as \mbox{e $\rightarrow$ e}. Both normal components of the wavevectors in free space and the anti-Hermitian medium are negative imaginary and we find the reflection factor for TE polarization as
\begin{equation}
    \label{eq:interface_r_ee}
    r_\mathrm{TE}^{\mathrm{e}\rightarrow\mathrm{e}} = \frac{\chi |k_{1x}|+\j |k_{2x}|}{\chi |k_{1x}| -\j |k_{2x}|} =\mathrm{e}^{\j\phi},
\end{equation}
which has unit magnitude.
As stated above, inside the anti-Hermitian medium propagating waves do not carry power, but evanescent waves do, which is opposite to the behavior in free space. Consequently, the evanescent wave in free space carries power along the $z$-direction (the propagation direction), whereas the evanescent wave in the medium carries power along the $x$-direction (the evanescent direction). For TE polarization, the time-averaged Poynting vector in the medium given by  Eq.~\eqref{eq:S_antihermitian_TE} points towards the interface in $-x$-direction for positive $\chi$ and it points away from the interface in $+x$-direction for negative $\chi$. This is because for positive (negative) $\chi$, the medium shows an overall gain (loss). To understand this result, we note that for a TE polarized evanescent wave the magnetic dissipated power is always larger than the electric dissipated power in the anti-Hermitian medium, $|\epsilon_0\chi|\mathbf{E}|^2|<|\mu_0\chi|\mathbf{H}|^2|$.  Thus, the magnetic response dominates, meaning that gain due to the permeability (positive $\chi$) leads to total gain in the anti-Hermitian medium, and loss in the permeability (negative $\chi$) leads to an overall loss. For a TM polarized evanescent wave, this conclusion is reversed: The electric dissipated power dominates, hence the permittivity determines whether the medium shows an overall gain or loss. 

Moreover, this is the fundamental reason why we need to retain only  evanescently decaying fields in the anti-Hermitian medium, even though it is an active medium and hence an exponentially growing solution is in principle allowed. To see why, let us consider TE polarization, for which the time-averaged Poynting vector is given by Eq.~\eqref{eq:S_antihermitian_TE}. The direction of the Poynting vector is determined by the sign of $\Im[k_{2x}]/\chi$. As discussed above, for positive $\chi$, the medium acts gainy and hence \emph{radiates power into free space}. The Poynting vector must therefore point in the negative $x$-direction, which enforces a negative sign of the imaginary part of $k_{2x}$. On the other hand, if $\chi$ is negative, the medium is lossy and \emph{absorbs power entering from free space}. Consequently, the Poynting vector must point in the positive $x$-direction, which again enforces a negative imaginary part of $k_{2x}$. This proves that there can only be evanescently decaying waves in the anti-Hermitian medium.

However, there is an apparent contradiction: Both incident and reflected waves are evanescent and hence, on their own, do not carry power along the $x$-direction. But inside the medium, depending on the sign of $\chi$, power is flowing in $+x$- or $-x$-direction, and due to the continuity of the $x$-component of the Poynting vector at the interface, there must be a nonzero $x$-component of the Poynging vector in free space. But how can power flow in the $x$-direction in free space? The answer lies in the \emph{phase} of the reflection factor. By superimposing forward and backward evanescent waves, the latter with a phase shift of $\phi$ as seen in Eq.~\eqref{eq:interface_r_ee}, the time-averaged Poynting vector shows a nonzero component along the evanescent decay direction proportional to $\sin\phi$. Recall that in the case \mbox{e $\rightarrow$ p} discussed above, the reflection factor is purely real, and hence the sine of the phase of the reflection factor is always zero, ensuring a zero time-averaged Poynting vector in the $x$-direction.

Numerical simulation results for $\chi=\pm2$ for TE polarization are shown in Fig.~\ref{fig:interface}(d). For negative $\chi$, the medium is lossy, and the Poynting vector inside the medium points away from the interface as the incident power is dissipated. In free space, the Poynting vector bends towards the interface. For positive $\chi$, the medium is gainy, and the Poynting vector points towards the interface in the medium. In free space, the Poynting vector bends away from the interface, as the medium radiates towards the $-x$-direction.


The case of a propagating incident wave has been discussed in some detail in the literature \cite{Xu:17_CCM, PhysRevA.96.043838_CPA_laser}. As briefly mentioned above, if the transmitted wave in the medium is propagating as well, the reflection factor is a phase factor of magnitude one, and there is no power transfer into the medium. Even so, there are reactive fields inside the medium, which can be used for e.g. sensing applications.

If the transmitted wave is evanescent, denoted as \mbox{p $\rightarrow$ e}, $k_{1x}$ is real and positive and $k_{2x} = -\mathrm{j}|k_{2x}|=-\mathrm{j}\sqrt{k_z^2-\chi^2k_0^2}$. This situation can take place only for $|\chi|<1$ and for angles of incidence larger than the critical angle of total internal reflection $\theta_\text{c}=\arcsin{(|\chi|)}$. The reflection factor Eq.~\eqref{eq:Interface_r_t} for TE polarization reads
\begin{equation}
    r_\text{TE}^{\mathrm{p}\rightarrow \mathrm{e}} = \frac{\chi k_{1x} +|k_{2x}|}{\chi k_{1x} - |k_{2x}|},\label{eq:interface_r_pe}
\end{equation}
and it is always larger than unity for positive $\chi$ ($\mu_\subr$ gainy) and smaller than unity for negative $\chi$ ($\mu_\subr$ lossy). As a result, the interface shows gain for positive $\chi$ and loss for negative $\chi$, the underlying reason of which was already discussed for the case \mbox{e $\rightarrow$ e}. Further, the time-averaged Poynting vector in Eq.~\eqref{eq:S_antihermitian_TE} is non-zero in the anti-Hermitian medium and points towards the interface in the negative $x$-direction for positive $\chi$ (the gainy medium radiates back to the free space), and away from the interface in the positive $x$-direction for negative $\chi$ (the lossy medium dissipates incident power). 
Setting the numerator in Eq.~\eqref{eq:interface_r_pe} to zero, we find the same condition for $k_z$ leading to infinite or zero reflection coefficient for positive or negative $\chi$ as in \eqref{eq:Interface_kz}, which can be converted into an angle of incidence given by $k_z = k_0\sin\theta_{1c}$.
 
It is worth emphasizing that for positive $\chi$, any infinitesimal incident field satisfying Eq.~\eqref{eq:Interface_kz} experiences infinite reflection, which makes this interface a laser without the need of a resonator cavity. 
For TM polarization, we again find zero or infinite reflection coefficients for the condition stated in Eq.~\eqref{eq:Interface_kz}, but for the opposite signs of $\chi$ as compared to TE polarization. As a result, for a plane wave incident under the critical angle $\theta_\text{1c}$, waves of one of the polarizations experience infinite reflection, while waves of the other experience zero reflection, which is an interesting generalization of the Brewster-angle phenomenon. 
One might raise a question of what happens when the incident wave  approaches normal incidence, i.e., $k_z\rightarrow 0$, because in that case there is no difference between TE  and TM polarizations. However, as soon as $k_z=0$, $k_{2x}$ is real for any value of $\chi$,  we are no longer in the propagating to evanescent regime and hence there is no lasing or perfect absorption possible.

Finally, the reflection factors for TE  and TM polarizations for the four different combinations of incident and  transmitted  waves are summarized in Table~\ref{tab:interface_summary}.

\begin{figure*}
\centering
\includegraphics[width=0.95\linewidth]{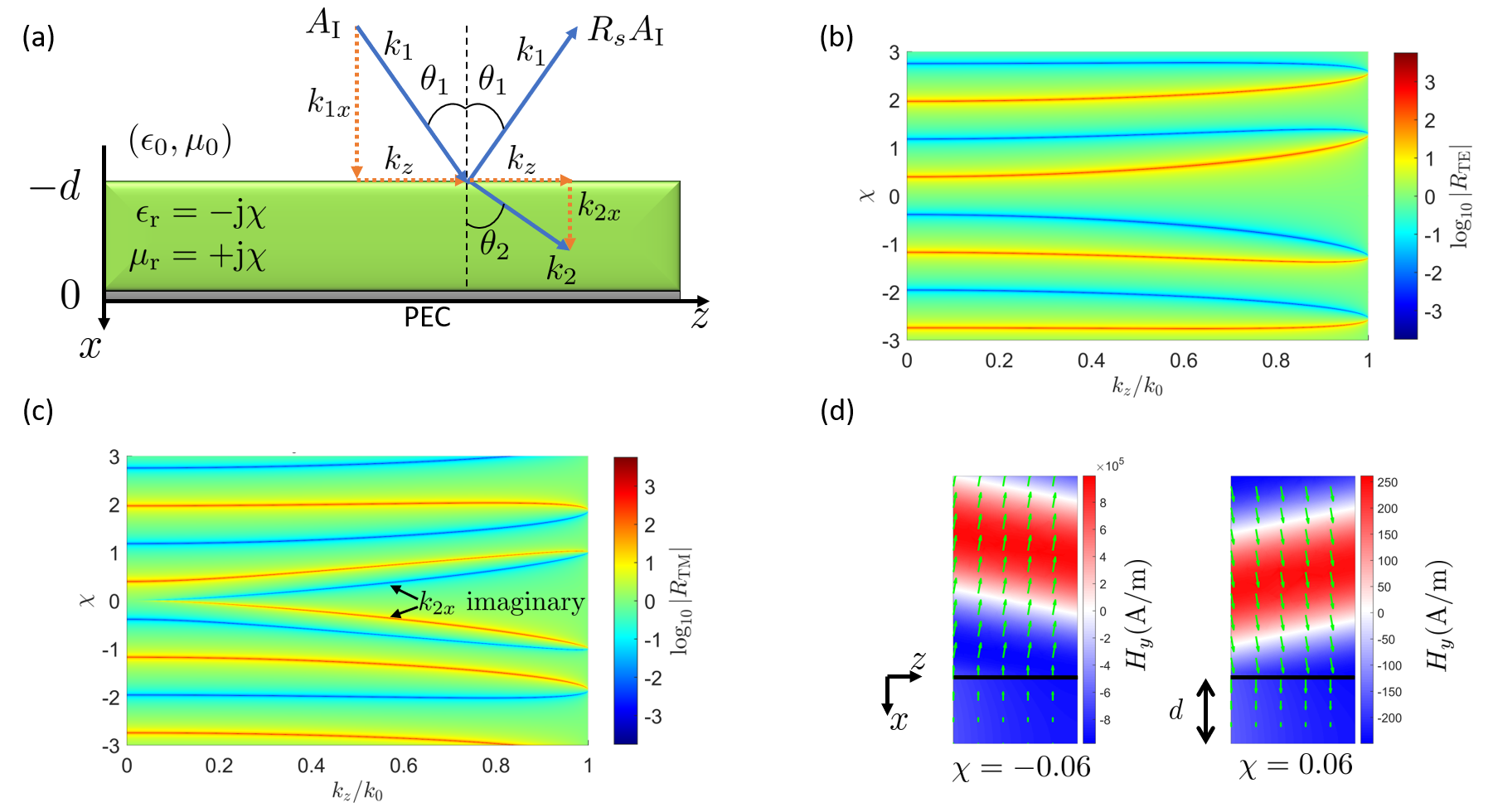}
\caption{ (a) An anti-Hermitian slab at a PEC boundary. An incident plane wave with amplitude $A_{\rm{I}}$ is reflected by the slab with reflection factor $R_s$, where $s=\text{TE,TM}$. (b) Absolute value of the reflection factor for TE illumination of a slab at a PEC boundary for $k_0d=2$ in logarithmic scale. One can clearly see the regions  where the slab acts as a laser (red) or as a perfect absorber (blue). (c) Same as (b) but for TM polarization. Interestingly, two additional regions for infinite and zero reflection appear for small values of $\chi$. These correspond to excitation of evanescent waves in the slab. (d) Simulation of an anti-Hermitian slab backed by a PEC boundary for TM polarization for an angle yielding infinite or zero reflection, depending on the sign of $\chi$. The slab thickness is $k_{0}d=2$, the incidence angle is $\theta_1=10.64^{\circ}$. The green arrows show the time-averaged Poynting vector.}
\label{fig:simulations}
\end{figure*}



\begin{table}
\caption{Summary of reflection factors for an interface between free space and an anti-Hermitian medium.}

\begin{tabular}{ |p{1.5cm}||c|c|  }
 \hline
 \textbf{Case} & \textbf{TE} & \textbf{TM}\\
 \hline
 \hline
  p $\rightarrow$ p & $|r_{\rm{TE}}|=1$ & $|r_{\rm{TM}}|=1$\\
 \hline
  e $\rightarrow$ e  & $|r_{\rm{TE}}|=1$ & $|r_{\rm{TM}}|=1$\\
 \hline
  p $\rightarrow$ e  & $|r_{\rm{TE}}|=\begin{cases}>1,\,\chi>0 \\ < 1,\,\chi<0\end{cases}$ &$|r_{\rm{TM}}|=\begin{cases}>1,\,\chi<0 \\ < 1,\,\chi>0\end{cases}$\\
 \hline
  e $\rightarrow$ p  &  $|r_{\rm{TE}}|=\begin{cases}>1,\,\chi<0 \\ < 1,\,\chi>0\end{cases}$ &$|r_{\rm{TM}}|=\begin{cases}>1,\,\chi>0 \\ < 1,\,\chi<0\end{cases}$\\
 \hline
\end{tabular}
\label{tab:interface_summary}
\end{table}



\subsection{Anti-Hermitian slabs backed by a mirror} 
Next, we consider a slab of thickness $d$, infinitely extended in $y$- and $z$-directions, filled by an anti-Hermitian medium. The volume $x<-d$ is free space, and there is a perfect electric conductor (PEC) boundary  at $x=0$, see Fig.~\ref{fig:simulations}(a). Let a TE polarized plane wave with  electric field along the $y$-axis be incident under an angle $\theta_1$ with respect to the surface normal of the slab. The electric field in the half-space above the slab $x<-d$ (region~I) is the superposition of incident and reflected waves $\mathbf{E}_\text{I}=A_\text{I}(\mathrm{e}^{-\mathrm{j}k_{1x}(x+d)}+R_\text{TE}\mathrm{e}^{\mathrm{j}k_{1x}(x+d)})\mathbf{\hat{a}}_y$, where $k_{1x}=\sqrt{k_0^2-k_z^2}$, $R_\mathrm{TE}$ is the  reflection factor of the grounded slab and $A_{\rm{I}}$ is the amplitude of the incident field. Similarly, the electric field inside the slab (region~II) is a superposition of forward and backward propagating plane waves $\mathbf{E}_\text{II}=A_\text{II}(\mathrm{e}^{-\mathrm{j}k_{2x}x}-\mathrm{e}^{\mathrm{j}k_{2x}x})\mathbf{\hat{a}}_y$, where the PEC boundary condition at $x=0$ is already satisfied, ${k_{2x} =\sqrt{\chi^2k_0^2-k_z^2}}$, and $A_{\rm{II}}$ is the amplitude of the fields inside the slab. For readability, the common propagator $\mathrm{e}^{-\mathrm{j}k_zz}$ is dropped. From Maxwell's curl equation $\nabla \times \mathbf{E} = -\mathrm{j}\omega\mu_0\mu_\subr\mathbf{H}$ we find the corresponding magnetic fields in the regions outside and inside the slab, and by imposing the boundary conditions at $x=-d$ we find the amplitude reflection factor
\begin{equation}
    R_\text{TE} = \frac{\chi k_{1x}\tan{(k_{2x}d)} + k_{2x}}{\chi k_{1x}\tan{(k_{2x}d)} - k_{2x}}. \label{eq:slab_R_PEC_TE}
\end{equation}
Following an analogous analysis for TM polarization, where we make a field {\it Ansatz} for the $y$-component of the magnetic field and calculate the electric field from $\nabla\times\mathbf{H}=\mathrm{j}\omega\epsilon_0\epsilon_\subr\mathbf{E}$, we find the amplitude reflection factor
\begin{equation}
    R_\text{TM} = \frac{\chi k_{1x}+ k_{2x}\tan{(k_{2x}d)}}{\chi k_{1x} - k_{2x}\tan{(k_{2x}d)}}. \label{eq:slab_R_PEC_TM}
\end{equation}
Comparing Eqs.~(\ref{eq:slab_R_PEC_TE} and \ref{eq:slab_R_PEC_TM}), it is evident that fields of TE  and TM polarizations behave  fundamentally differently in this system. The simple  conversion rule between polarization states by changing the sign of $\chi$ as in the interface problem is no longer valid, which is because the symmetry of TE  and TM polarizations is broken by the PEC boundary.
If the incident wave is evanescent ($k_{1x}$ imaginary), the reflection factor for both TE and TM polarization always has magnitude one, which can be easily verified from Eqs.~(\ref{eq:slab_R_PEC_TE} and \ref{eq:slab_R_PEC_TM}). If the incident wave is propagating, the reflection factor for both TE  and TM polarizations can in principle diverge (lasing) or go to zero (perfect absorption) for both propagating and evanescent waves in the slab. Figures~\ref{fig:simulations}(b) and (c) show the absolute values of the reflection factor for TE   and TM polarizations, respectively, for $0\leq k_z/k_0\leq1$, $-3\leq\chi\leq3$,  and the slab thickness $k_0d=2$.

One key result is the fact that for the TM polarization lasing can be achieved for an arbitrarily small value of $\chi$ at near normal incidence, as long as $\chi$ is negative. This branch in Fig.~\ref{fig:simulations}(c) corresponds to evanescent waves inside the slab, as $|\chi|< k_z/k_0$. As discussed for the interface above, TM polarized evanescent waves are dominated by the electric response, which is again confirmed by the fact that within these additional branches for the TM polarization, the slab shows gain for negative $\chi$ and loss for positive $\chi$. 

For grazing incidence $k_z\rightarrow k_0$, there are exotic points in the reflection factors for specific values of $\chi$, where the infinite and zero reflection curves intersect. To understand this effect, let us see when the reflection factor for TE polarization Eq.~\eqref{eq:slab_R_PEC_TE} has a null or a pole in the case of $k_z\rightarrow k_0$. We need to set the numerator or denominator in Eq.~\eqref{eq:slab_R_PEC_TE} to zero, resulting in the condition
\begin{equation}
\label{eq:Slab_R_grazing}
    \chi\sqrt{k_0^2-k_z^2}\tan\left(d\sqrt{\chi^2k_0^2-k_z^2}\right) = \pm\sqrt{\chi^2k_0-k_z^2}
\end{equation}
where the plus sign is for a pole and the minus sign is for a null. If we assume $\chi$ to be finite and not equal to one, we see that the right-hand side of Eq.~\eqref{eq:Slab_R_grazing} is finite. Since on the left-hand side the term $\sqrt{k_0^2-k_z^2}$ tends to zero, we conclude that the tangent term must diverge, i.e., $d\sqrt{\chi^2k_0^2-k_z^2} = \pi(1/2+m),\,m\in\mathbb{Z}$. For $k_z\rightarrow k_0$ this condition simplifies to $k_0d\sqrt{\chi^2-1} = \pi(1/2+m)$. Note that the tangent term tends to both plus and minus infinity at these points, which means that they simultaneously correspond to nulls and poles, according to Eq.~\eqref{eq:Slab_R_grazing}. For $m=0$ and $k_0d=2$ we find $\chi=\sqrt{1+(\pi/4)^2}\approx 1.27$, which agrees with Fig.~\ref{fig:simulations}(b). For TM polarization, we find the analogous condition $k_0d\sqrt{\chi^2-1}=m\pi.$

Simulation results for a TM polarized plane wave incident under an angle of $\theta_1=10.64^\circ$ on the slab for $k_0d=2$ and $\chi=\pm0.06$ are shown in Fig.~\ref{fig:simulations}(d). For $\chi=-0.06$, the slab acts as a laser, for $\chi=0.06$ the slab acts as a perfect absorber. The numerical values for the reflection factor in simulations agree with the theoretical results calculated from Eq.~\eqref{eq:slab_R_PEC_TM}. 


\subsection{Anti-Hermitian metasurfaces}
Let us next discuss illumination of a PEC-backed layer by a normally incident wave with $k_z=0$, meaning that $k_{1x}=k_0$ and $k_{2x}=|\chi| k_0$. We find the reflection factors from Eqs.~(\ref{eq:slab_R_PEC_TE} and \ref{eq:slab_R_PEC_TM}) as
\begin{align}
\label{eq:slab_R_PEC_normal}
\begin{split}
    R_\text{TE}=\frac{\tan{\delta}+\text{sgn}[\chi]}{\tan{\delta}-\text{sgn}[\chi]},\quad    R_\text{TM} = -R_\text{TE},
\end{split}
\end{align}
where $\text{sgn}[\chi]=|\chi|/\chi$ and $\delta=k_0d|\chi|$. For positive values of $\chi$, the reflection factors diverge when $\delta=(4m+1)\pi/4$, where $m=0,1,2$\dots. Accordingly, by assuming that $\chi\gg 1$, the divergence of both $R_\text{TE}$ and $R_\text{TM}$ can take place for $m=0$ resulting in $k_0d=(\pi/4\chi)\ll 1$, which means that the slab is electrically very thin as compared to $\lambda=2\pi/k_0$. If the slab is located in free space, similar divergence of $R_\text{TE}$ and $R_\text{TM}$ holds at resonances, and the corresponding value $k_0d=(\pi/2\chi)$ \cite{PhysRevA.96.043838_CPA_laser} can be also smaller than unity if $\chi\gg 1$. This result indicates that one can engineer creation of a normally radiated wave from an anti-Hermitian metasurface. 

However, it is not possible to model metasurfaces (sheets of negligible thickness) with volumetric material parameters, as in Eq.~\eqref{eq:slab_R_PEC_normal}. To properly describe lasing from anti-Hermitian metasurfaces formed by single-layer arrays of small particles, we use the surface susceptibility model, e.g. \cite{Glybovski}. In our case of a thin magnetodielectric layer, the electric and magnetic susceptibilities of the metasurface  can be defined in a scalar form, relating the surface densities of polarizations with the tangential fields: $\chi_{ee}=P_{s}/\langle E\rangle$ and $\chi_{mm}=M_{s}/\langle H\rangle$. Here, brackets denote averaging of macroscopic $E$ and $H$ fields taken on both sides of the metasurface. 
Expressing the susceptibilities in terms of the material parameters of the slab, we find that for an anti-Hermitian metasurface  $\chi_{mm}=\pm \j\xi=\mp \chi_{ee}\eta_0^2$. The reflection coefficient of such a metasurface is expressed via $\chi_{ee}$ and $\chi_{mm}$ in Eq.~(4) of Ref.~\cite{Glybovski}. Equating the denominator of that expression (in which we should nullify the bianisotropic susceptibilities) to zero, we easily obtain the effective parameters of the anti-Hermitian metasurface as
\begin{equation}
\chi_{mm}=\frac{2\eta_0}{\mathrm{j}\omega}, \quad \chi_{ee} =-\frac{2}{\mathrm{j} \eta_0\omega},
\label{eq:MS_chi}
\end{equation}
which grant lasing from the metasurface in the normal direction. In practice, an anti-Hermitian metasurface can be realized as a dense array of small resonant electric and magnetic scatterers. The electric or magnetic gain in them can be offered by some pumping, whereas purely lossy magnetic or electric response arises at the Lorentzian resonance naturally.  

We have hitherto discussed how to realize anti-Hermitian metasurface lasing in broadside direction. However, one can generalize this theory also for oblique incidence and find poles of the reflection coefficients also for other angles. Concerning this scenario, first, by equating the denominators in Eqs.~\eqref{eq:slab_R_PEC_TE} and \eqref{eq:slab_R_PEC_TM} to zero, we find that it is indeed possible for both reflection factors to diverge under the conditions $k_0d\ll1$ and $\chi\gg1$. This suggests that there is a possibility to realize this response by a metasurface with proper values of  susceptibilities. We need to only calculate the analogous expressions to Eq.~\eqref{eq:MS_chi} to find the metasurface parameters that provide lasing effect at a particular angle.


\section{Mie Scattering from an anti-Hermitian Spherical Particle} 
\label{sec:GSpheG}

\begin{figure*}[t]
\centering
\includegraphics[width = 0.95\linewidth]{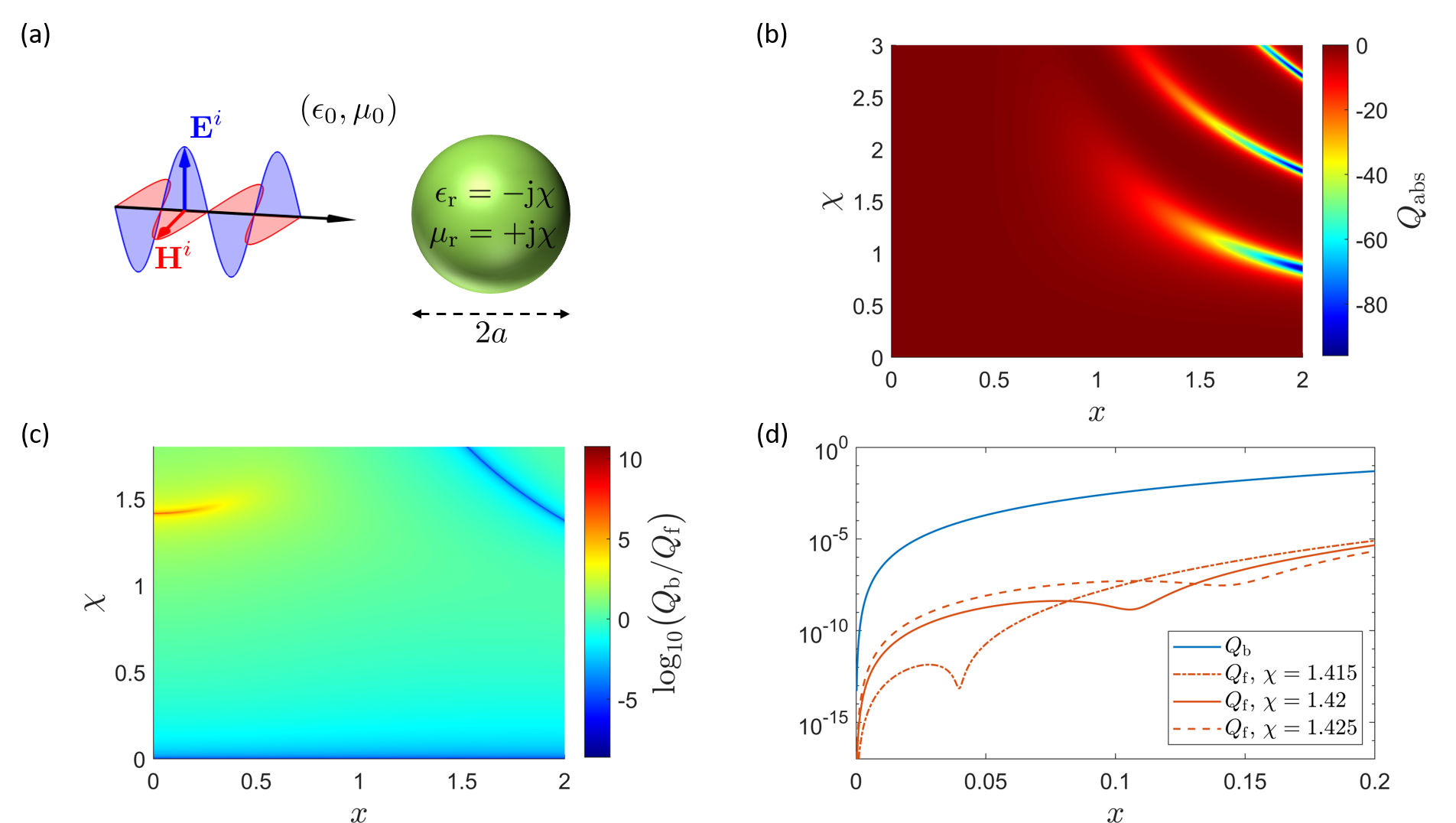}
\caption{(a) Plane-wave scattering by an anti-Hermitian sphere. (b) The absorption efficiency $Q_\text{abs}$ of an anti-Hermitian sphere as a function of both the material parameter $0<\chi<3$ and the size parameter $x$ up to the value of $2$. (c) The back-to-forward scattering cross section ratio of an anti-Hermitian sphere as a function of its size parameter $x$ and the material parameter $\chi$. There is a singularity for $\chi=\sqrt{2}$ as the sphere size approaches zero. (d) Forward and backward scattering efficiencies as a function of the sphere size in the dipolar limit for three different values of $\chi$, which are close to $\sqrt2$. The forward scattering efficiency is very sensitive to changes in $\chi$, whereas the backward scattering efficiency is approximately constant.}
\label{fig:sphere}
\end{figure*}

The interaction of light with objects having spherical symmetry, as depicted in Fig.~\ref{fig:sphere}(a), can be analyzed using the classical Mie scattering principles~\cite{Bohren-Huffman, Ari_Mie, Osipov_book}. Measures for the effect of the particle on the incoming radiation are its scattering, absorption, and extinction cross section. Depending on the relative permittivity $\epsilon_\subr$ and relative permeability $\mu_\subr$ of the material from which the sphere is made, and, also, the optical size parameter $x=2\pi a/\lambda$ ($\lambda$ is the wavelength in free space, and $a$ represents the radius), the three normalized efficiencies read 
\begin{equation}
\begin{split}
&Q_\text{sca}=\frac{C_\text{sca}}{\pi a^2}=\frac{2}{x^2} \sum_{n=1}^{\infty }(2n+1)\left(\left| a_n \right|^2 + \left| b_n \right|^2\right),\cr
&Q_\text{ext}=\frac{C_\text{ext}}{\pi a^2}=\frac{2}{x^2}{\sum_{n=1}^{\infty }}(2n+1)\operatorname{Re}\left[a_n + b_n\right],
\end{split}
\end{equation}
and $Q_\text{abs}=Q_\text{ext}-Q_\text{sca}$. Here, $C_\text{sca}$ and $C_\text{ext}$ are the scattering and extinction cross sections, respectively. The electric and magnetic Mie coefficients appearing in these expressions are given as functions of the primary parameters of the sphere by 
\begin{equation}
\label{eq:anbn}
\begin{aligned}
a_n &= \frac{\sqrt{\epsilon_\subr}\psi_n(\sqrt{\epsilon_\subr\mu_\subr}\,x)\psi_n'(x) 
- \sqrt{\mu_\subr}\,\psi_n(x)\psi_n'(\sqrt{\epsilon_\subr\mu_\subr}\,x)}
{\sqrt{\epsilon_\subr}\psi_n(\sqrt{\epsilon_\subr\mu_\subr}\,x)\xi_n'(x) 
- \sqrt{\mu_\subr}\,\xi_n(x)\psi_n'(\sqrt{\epsilon_\subr\mu_\subr}\,x)},\\[1ex]
b_n &= \frac{\sqrt{\mu_\subr}\psi_n(\sqrt{\epsilon_\subr\mu_\subr}\,x)\psi_n'(x) 
- \sqrt{\epsilon_\subr}\,\psi_n(x)\psi_n'(\sqrt{\epsilon_\subr\mu_\subr}\,x)}
{\sqrt{\mu_\subr}\psi_n(\sqrt{\epsilon_\subr\mu_\subr}\,x)\xi_n'(x) 
- \sqrt{\epsilon_\subr}\,\xi_n(x)\psi_n'(\sqrt{\epsilon_\subr\mu_\subr}\,x)}.
\end{aligned}
\end{equation}
The Riccati--Bessel functions $\psi_n$ and $ \xi_n$ are expressed as
\begin{equation}
\psi_n(\rho) = \rho\,j_n(\rho), \qquad
\xi_n(\rho) = \rho\,h_n^{(2)}(\rho), 
\end{equation}
by using the ordinary spherical Bessel ($j_n$) and Hankel ($h_n$) functions. It is worth noticing that in the electrical engineering convention that we use in this paper, the Hankel functions of second kind with an asymptotic form $\exp(-\j\rho)/\rho$ as $\rho\rightarrow\infty$ must be employed for outgoing spherical waves. 

Based on these formulas, Fig.~\ref{fig:sphere}(b) illustrates the Mie scattering theory prediction for the absorption efficiency $Q_\text{abs}$ of an anti-Hermitian sphere in the parameter space ($\chi<3$ and $x<2$). Remarkably, the absorption efficiency is negative for all values of $x$ and $\chi$, meaning that the sphere always exhibits gainy behaviour. 
Thus, although the material is magneto-electrically symmetric in terms of gain and loss, a finite object displays a global gain.
Moreover, the sphere treated in Fig.~\ref{fig:sphere}(b) is characterized by loss in the permittivity and gain in the permeability. However, there exists a true symmetry: Recomputing the absorption efficiency for a dual sphere where permittivity is gainy and permeability lossy with the same amplitudes returns us exactly the same plot as in Figure~\ref{fig:sphere}(b). This is understandable considering that the electric $(a_n)$ and magnetic $(b_n)$ Mie coefficients (which add up equally in the efficiency expansions) are dual in the sense that they exchange values when permittivity and permeability are swapped: $a_n(\epsilon_\subr,\mu_\subr)=b_n(\mu_\subr,\epsilon_\subr)$, as is clear from Eq.~\eqref{eq:anbn}.

Let us investigate how the anti-Hermitian characteristic affects the backward to forward scattering ratio $Q_\text{b}/Q_\text{f}$ for optically small particles. At small values of $x$, we use  the dipole approximation, retaining only the terms $n=1$. By employing the Taylor expansion of spherical Bessel and Hankel functions, we infer that $a_1=-(2x^3/3\j)(\epsilon_\subr-1)/(\epsilon_\subr+2)$ and $b_1=-(2x^3/3\j)(\mu_\subr-1)/(\mu_\subr+2)$. If we substitute these expressions into the ratio given by $Q_{\rm{b}}/Q_{\rm{f}}=\vert a_1-b_1\vert^2/\vert a_1+b_1\vert^2$, we find that
\begin{equation}
    \frac{Q_{\rm{b}}}{Q_{\rm{f}}}=\frac{9\vert\epsilon_\subr-\mu_\subr\vert^2}{\vert\epsilon_\subr+\mu_\subr+2\epsilon_\subr\mu_\subr-4\vert^2},
    \label{eq:Qb:Qf}
\end{equation}
see, e.g., Ref.~\cite{Osipov_book}.
This relation is general, and it is not limited to the anti-Hermitian sphere. Obviously, backward scattering vanishes if $\epsilon_\subr=\mu_\subr$. Forward scattering vanishes when  the denominator is zero, which takes place when 
\begin{equation}
\epsilon_\subr=\frac{4-\mu_\subr}{1+2\mu_\subr}\qquad \text{or}\quad \mu_\subr=\frac{4-\epsilon_\subr}{1+2\epsilon_\subr}.
\label{eq:conemuw}
\end{equation} 
This is the so-called second Kerker condition \cite{Kerker83}. For passive spheres, this condition can be satisfied only for lossless particles \cite{Osipov_book}, and the apparent contradiction with the optical theorem (zero forward scattering while the extinction cross section is not zero) is resolved by accounting for higher-order terms in the Mie expansion \cite{Alu_Engheta_2010}.

If $\mu_\subr=1$, then $\epsilon_\subr=1$ which means that for dipolar non-magnetic spheres zero forward scattering is not possible even for active particles. It was shown that active dielectric  scatterers usually provide higher values of $Q_{\rm{b}}/Q_{\rm{f}}$, but this occurs for only larger values of $x$~\cite{ARISihvola2021EBAP}. However, using Eq.~\eqref{eq:conemuw}, we readily prove that if $\mu_\subr$ and $\epsilon_\subr$ are purely imaginary, the second Kerker condition is satisfied at $\epsilon_\subr=-\mu_\subr=\pm\j\sqrt{2}$. It is expected that this peculiar value results in having tremendous amount of backward scattering compared to the forward scattering. To confirm this theoretical derivation, we made corresponding full-wave simulations and plotted $Q_{\rm{b}}/Q_{\rm{f}}$ with respect to $\chi$ and $x$. As it is seen in Fig.~\ref{fig:sphere}(c), near $\chi\approx\sqrt{2}$, the ratio is indeed very large. This is because the forward scattering efficiency approaches zero, while the backward scattering efficiency is finite. In Fig.~\ref{fig:sphere}(d), the forward and backward scattering efficiencies for a small sphere ($x<0.2)$ and several values of $\chi$ being close to $\sqrt{2}$ are shown separately. We clearly observe that the forward scattering efficiency has a local minimum that shifts considerably to lower values of $x$ as $\chi$ is slightly decreasing and becoming close to $\sqrt{2}$, while the backward scattering efficiency is approximately insensitive to those changes in the value of $\chi$. Thus, firstly, this local minimum for the forward scattering efficiency gives rise to the large backward to forward scattering ratio, and, secondly, the forward scattering efficiency reaches zero only at exactly $\chi=\sqrt2$ for $x\rightarrow0$. For comparison, a conventional dielectric sphere with a refractive index of $\sqrt2$ shows approximately equal amounts of forward and backward scattering in the dipolar limit. This is in stark contrast to the anti-Hermitian sphere, where the forward scattering is strongly suppressed.

Another interesting feature in Fig.~\ref{fig:sphere}(c) is the fact that the backward to forward scattering ratio approaches zero for $\chi\rightarrow 0$, for any value of $x$. This is because the backward scattering efficiency tends to zero in this scenario, which can be explained by the fact that the relative permittivity and relative permeability approach each other. Then, from Eq.~\eqref{eq:anbn} it is clear that $a_n$ and $b_n$ approach each other, resulting in zero backscattering.
Lastly, for higher values of $x$ and $\chi$, there is a region in Fig.~\ref{fig:sphere}(c) where the backward to forward scattering ratio is close to zero, which is because the backward scattering efficiency shows a local minimum there.


\section{Conclusions} 
\label{sec:conslW}
This work contemplates electromagnetic effects in  anti-Hermitian objects with planar and spherical geometries. We have shown the behaviour of both evanescent and propagating plane waves incident on an interface between free space and an anti-Hermitian medium. If incident and transmitted waves are of the same kind (both evanescent or both propagating), the reflection factor is of magnitude one. However, if the a propagating wave is transformed into an evanescent wave or vice versa, both lasing and perfect absorption can occur, depending on the polarization and the sign of the anti-Hermitian parameter. In the lasing case, the lasing mode in free space is an evanescent (propagating) wave for an evanescent (propagating) incident wave. This phenomenon allows for a novel type of thin-layer laser, if the anti-Hermitian medium slab is terminated by an impedance boundary matched to the wave impedance. Of special interest is a possibility of a thin anti-Hermitian sheet creating an evanescent surface lasing mode that is tightly bound to the interface.

Another interesting scenario is an anti-Hermitian slab backed by a PEC mirror. We have found that such a slab can act both as a laser and perfect absorber, depending on the slab thickness, polarization of the incident wave, and sign of the anti-Hermitian parameter. For TM polarization, lasing and absorption can seemingly occur at arbitrarily low values of the anti-Hermitian parameter, which is of paramount practical interest. Furthermore, the anti-Hermitian slab can be reduced to a metasurface, allowing for lasing or perfect absorption at an arbitrarily thin anti-Hermitian sheet.

Finally, scattering of light by an anti-Hermitian sphere using Mie theory was studied. Within the parameter range analyzed here, the sphere always exhibits gain. This means that, even though microscopically the gain and loss in the medium are balanced, a macroscopic object like a sphere shows an overall gain. Moreover, for small sphere sizes, the backward to forward scattering ratio diverges for $\chi$ close to $\sqrt2$, which is due to the forward scattering exhibiting a local minimum. Compared to a conventional small dielectric sphere with refractive index of $\sqrt2$, the forward scattering in an anti-Hermitian sphere is strongly suppressed.


\section*{Acknowledgement} 
L.F. and M.S.M. thank Xuchen Wang, Markus Nyman, and Benedikt Zerulla for their invaluable help with numerical simulations. Also, L.F. wishes to acknowledge support of Carsten Rockstuhl.


\bibliography{sample}

\begin{thebibliography}{32}%
\makeatletter
\providecommand \@ifxundefined [1]{%
 \@ifx{#1\undefined}
}%
\providecommand \@ifnum [1]{%
 \ifnum #1\expandafter \@firstoftwo
 \else \expandafter \@secondoftwo
 \fi
}%
\providecommand \@ifx [1]{%
 \ifx #1\expandafter \@firstoftwo
 \else \expandafter \@secondoftwo
 \fi
}%
\providecommand \natexlab [1]{#1}%
\providecommand \enquote  [1]{``#1''}%
\providecommand \bibnamefont  [1]{#1}%
\providecommand \bibfnamefont [1]{#1}%
\providecommand \citenamefont [1]{#1}%
\providecommand \href@noop [0]{\@secondoftwo}%
\providecommand \href [0]{\begingroup \@sanitize@url \@href}%
\providecommand \@href[1]{\@@startlink{#1}\@@href}%
\providecommand \@@href[1]{\endgroup#1\@@endlink}%
\providecommand \@sanitize@url [0]{\catcode `\\12\catcode `\$12\catcode
  `\&12\catcode `\#12\catcode `\^12\catcode `\_12\catcode `\%12\relax}%
\providecommand \@@startlink[1]{}%
\providecommand \@@endlink[0]{}%
\providecommand \url  [0]{\begingroup\@sanitize@url \@url }%
\providecommand \@url [1]{\endgroup\@href {#1}{\urlprefix }}%
\providecommand \urlprefix  [0]{URL }%
\providecommand \Eprint [0]{\href }%
\providecommand \doibase [0]{https://doi.org/}%
\providecommand \selectlanguage [0]{\@gobble}%
\providecommand \bibinfo  [0]{\@secondoftwo}%
\providecommand \bibfield  [0]{\@secondoftwo}%
\providecommand \translation [1]{[#1]}%
\providecommand \BibitemOpen [0]{}%
\providecommand \bibitemStop [0]{}%
\providecommand \bibitemNoStop [0]{.\EOS\space}%
\providecommand \EOS [0]{\spacefactor3000\relax}%
\providecommand \BibitemShut  [1]{\csname bibitem#1\endcsname}%
\let\auto@bib@innerbib\@empty
\bibitem [{\citenamefont {Ashida}\ \emph {et~al.}(2020)\citenamefont {Ashida},
  \citenamefont {Gong},\ and\ \citenamefont {Ueda}}]{ashida2020non}%
  \BibitemOpen
  \bibfield  {author} {\bibinfo {author} {\bibfnamefont {Y.}~\bibnamefont
  {Ashida}}, \bibinfo {author} {\bibfnamefont {Z.}~\bibnamefont {Gong}},\ and\
  \bibinfo {author} {\bibfnamefont {M.}~\bibnamefont {Ueda}},\ }\bibfield
  {title} {\bibinfo {title} {{Non-Hermitian physics}},\ }\href
  {https://doi.org/10.1080/00018732.2021.1876991} {\bibfield  {journal}
  {\bibinfo  {journal} {Advances in Physics}\ }\textbf {\bibinfo {volume}
  {69}},\ \bibinfo {pages} {249} (\bibinfo {year} {2020})}\BibitemShut
  {NoStop}%
\bibitem [{\citenamefont {El-Ganainy}\ \emph {et~al.}(2018)\citenamefont
  {El-Ganainy}, \citenamefont {Makris}, \citenamefont {Khajavikhan},
  \citenamefont {Musslimani}, \citenamefont {Rotter},\ and\ \citenamefont
  {Christodoulides}}]{El2018nonHERM}%
  \BibitemOpen
  \bibfield  {author} {\bibinfo {author} {\bibfnamefont {R.}~\bibnamefont
  {El-Ganainy}}, \bibinfo {author} {\bibfnamefont {K.~G.}\ \bibnamefont
  {Makris}}, \bibinfo {author} {\bibfnamefont {M.}~\bibnamefont {Khajavikhan}},
  \bibinfo {author} {\bibfnamefont {Z.~H.}\ \bibnamefont {Musslimani}},
  \bibinfo {author} {\bibfnamefont {S.}~\bibnamefont {Rotter}},\ and\ \bibinfo
  {author} {\bibfnamefont {D.~N.}\ \bibnamefont {Christodoulides}},\ }\bibfield
   {title} {\bibinfo {title} {{Non-Hermitian physics and PT symmetry}},\ }\href
  {https://doi.org/10.1038/NPHYS4323} {\bibfield  {journal} {\bibinfo
  {journal} {Nature Phys}\ }\textbf {\bibinfo {volume} {14}},\ \bibinfo {pages}
  {11} (\bibinfo {year} {2018})}\BibitemShut {NoStop}%
\bibitem [{\citenamefont {Feng}\ \emph {et~al.}(2017)\citenamefont {Feng},
  \citenamefont {El-Ganainy},\ and\ \citenamefont {Ge}}]{feng2017NHBOPTSN}%
  \BibitemOpen
  \bibfield  {author} {\bibinfo {author} {\bibfnamefont {L.}~\bibnamefont
  {Feng}}, \bibinfo {author} {\bibfnamefont {R.}~\bibnamefont {El-Ganainy}},\
  and\ \bibinfo {author} {\bibfnamefont {L.}~\bibnamefont {Ge}},\ }\bibfield
  {title} {\bibinfo {title} {{Non-Hermitian photonics based on parity-–time
  symmetry}},\ }\href {https://doi.org/10.1038/s41566-017-0031-1} {\bibfield
  {journal} {\bibinfo  {journal} {Nature Photonics}\ }\textbf {\bibinfo
  {volume} {11}},\ \bibinfo {pages} {752} (\bibinfo {year} {2017})}\BibitemShut
  {NoStop}%
\bibitem [{\citenamefont {Longhi}(2017)}]{Longhi2018PTRS}%
  \BibitemOpen
  \bibfield  {author} {\bibinfo {author} {\bibfnamefont {S.}~\bibnamefont
  {Longhi}},\ }\bibfield  {title} {\bibinfo {title} {{Parity-time symmetry
  meets photonics: A new twist in non-Hermitian optics}},\ }\href
  {https://doi.org/10.1209/0295-5075/120/64001} {\bibfield  {journal} {\bibinfo
   {journal} {Europhysics Letters}\ }\textbf {\bibinfo {volume} {120}},\
  \bibinfo {pages} {64001} (\bibinfo {year} {2017})}\BibitemShut {NoStop}%
\bibitem [{\citenamefont {Wong}\ \emph {et~al.}(2016)\citenamefont {Wong},
  \citenamefont {Xu}, \citenamefont {Kim}, \citenamefont {O'Brien},
  \citenamefont {Wang}, \citenamefont {Feng},\ and\ \citenamefont
  {Zhang}}]{Wong2016lasingNL}%
  \BibitemOpen
  \bibfield  {author} {\bibinfo {author} {\bibfnamefont {Z.~J.}\ \bibnamefont
  {Wong}}, \bibinfo {author} {\bibfnamefont {Y.-L.}\ \bibnamefont {Xu}},
  \bibinfo {author} {\bibfnamefont {J.}~\bibnamefont {Kim}}, \bibinfo {author}
  {\bibfnamefont {K.}~\bibnamefont {O'Brien}}, \bibinfo {author} {\bibfnamefont
  {Y.}~\bibnamefont {Wang}}, \bibinfo {author} {\bibfnamefont {L.}~\bibnamefont
  {Feng}},\ and\ \bibinfo {author} {\bibfnamefont {X.}~\bibnamefont {Zhang}},\
  }\bibfield  {title} {\bibinfo {title} {{Lasing and anti-lasing in a single
  cavity}},\ }\href {https://doi.org/10.1038/nphoton.2016.216} {\bibfield
  {journal} {\bibinfo  {journal} {Nature Photonics}\ }\textbf {\bibinfo
  {volume} {10}},\ \bibinfo {pages} {796} (\bibinfo {year} {2016})}\BibitemShut
  {NoStop}%
\bibitem [{\citenamefont {Longhi}(2010)}]{longhi2010ptLAL}%
  \BibitemOpen
  \bibfield  {author} {\bibinfo {author} {\bibfnamefont {S.}~\bibnamefont
  {Longhi}},\ }\bibfield  {title} {\bibinfo {title} {$\mathcal{PT}$-symmetric
  laser absorber},\ }\href {https://doi.org/10.1103/PhysRevA.82.031801}
  {\bibfield  {journal} {\bibinfo  {journal} {Phys. Rev. A}\ }\textbf {\bibinfo
  {volume} {82}},\ \bibinfo {pages} {031801} (\bibinfo {year}
  {2010})}\BibitemShut {NoStop}%
\bibitem [{\citenamefont {Chong}\ \emph {et~al.}(2011)\citenamefont {Chong},
  \citenamefont {Ge},\ and\ \citenamefont {Stone}}]{Chong2011_PTLaserAbsorber}%
  \BibitemOpen
  \bibfield  {author} {\bibinfo {author} {\bibfnamefont {Y.~D.}\ \bibnamefont
  {Chong}}, \bibinfo {author} {\bibfnamefont {L.}~\bibnamefont {Ge}},\ and\
  \bibinfo {author} {\bibfnamefont {A.~D.}\ \bibnamefont {Stone}},\ }\bibfield
  {title} {\bibinfo {title} {{$\mathcal{P}\mathcal{T}$-Symmetry Breaking and
  Laser-Absorber Modes in Optical Scattering Systems}},\ }\href
  {https://doi.org/10.1103/PhysRevLett.106.093902} {\bibfield  {journal}
  {\bibinfo  {journal} {Phys. Rev. Lett.}\ }\textbf {\bibinfo {volume} {106}},\
  \bibinfo {pages} {093902} (\bibinfo {year} {2011})}\BibitemShut {NoStop}%
\bibitem [{\citenamefont {Hodaei}\ \emph {et~al.}(2014)\citenamefont {Hodaei},
  \citenamefont {Miri}, \citenamefont {Heinrich}, \citenamefont
  {Christodoulides},\ and\ \citenamefont {Khajavikhan}}]{Hodaei2014SMLE}%
  \BibitemOpen
  \bibfield  {author} {\bibinfo {author} {\bibfnamefont {H.}~\bibnamefont
  {Hodaei}}, \bibinfo {author} {\bibfnamefont {M.-A.}\ \bibnamefont {Miri}},
  \bibinfo {author} {\bibfnamefont {M.}~\bibnamefont {Heinrich}}, \bibinfo
  {author} {\bibfnamefont {D.~N.}\ \bibnamefont {Christodoulides}},\ and\
  \bibinfo {author} {\bibfnamefont {M.}~\bibnamefont {Khajavikhan}},\
  }\bibfield  {title} {\bibinfo {title} {Parity-time--symmetric microring
  lasers},\ }\href {https://doi.org/10.1126/SCIENCE.1258480} {\bibfield
  {journal} {\bibinfo  {journal} {Science}\ }\textbf {\bibinfo {volume}
  {346}},\ \bibinfo {pages} {975} (\bibinfo {year} {2014})}\BibitemShut
  {NoStop}%
\bibitem [{\citenamefont {Feng}\ \emph {et~al.}(2014)\citenamefont {Feng},
  \citenamefont {Wong}, \citenamefont {Ma}, \citenamefont {Wang},\ and\
  \citenamefont {Zhang}}]{feng2014SMLPTSB}%
  \BibitemOpen
  \bibfield  {author} {\bibinfo {author} {\bibfnamefont {L.}~\bibnamefont
  {Feng}}, \bibinfo {author} {\bibfnamefont {Z.~J.}\ \bibnamefont {Wong}},
  \bibinfo {author} {\bibfnamefont {R.-M.}\ \bibnamefont {Ma}}, \bibinfo
  {author} {\bibfnamefont {Y.}~\bibnamefont {Wang}},\ and\ \bibinfo {author}
  {\bibfnamefont {X.}~\bibnamefont {Zhang}},\ }\bibfield  {title} {\bibinfo
  {title} {{Single-mode laser by parity-time symmetry breaking}},\ }\href
  {https://doi.org/10.1126/science.1258479} {\bibfield  {journal} {\bibinfo
  {journal} {Science}\ }\textbf {\bibinfo {volume} {346}},\ \bibinfo {pages}
  {972} (\bibinfo {year} {2014})}\BibitemShut {NoStop}%
\bibitem [{\citenamefont {Miao}\ \emph {et~al.}(2016)\citenamefont {Miao},
  \citenamefont {Zhang}, \citenamefont {Sun}, \citenamefont {Walasik},
  \citenamefont {Longhi}, \citenamefont {Litchinitser},\ and\ \citenamefont
  {Feng}}]{miao2016OAML}%
  \BibitemOpen
  \bibfield  {author} {\bibinfo {author} {\bibfnamefont {P.}~\bibnamefont
  {Miao}}, \bibinfo {author} {\bibfnamefont {Z.}~\bibnamefont {Zhang}},
  \bibinfo {author} {\bibfnamefont {J.}~\bibnamefont {Sun}}, \bibinfo {author}
  {\bibfnamefont {W.}~\bibnamefont {Walasik}}, \bibinfo {author} {\bibfnamefont
  {S.}~\bibnamefont {Longhi}}, \bibinfo {author} {\bibfnamefont {N.~M.}\
  \bibnamefont {Litchinitser}},\ and\ \bibinfo {author} {\bibfnamefont
  {L.}~\bibnamefont {Feng}},\ }\bibfield  {title} {\bibinfo {title} {Orbital
  angular momentum microlaser},\ }\href
  {https://doi.org/10.1126/SCIENCE.AAF8533} {\bibfield  {journal} {\bibinfo
  {journal} {Science}\ }\textbf {\bibinfo {volume} {353}},\ \bibinfo {pages}
  {464} (\bibinfo {year} {2016})}\BibitemShut {NoStop}%
\bibitem [{\citenamefont {Lin}\ \emph {et~al.}(2011)\citenamefont {Lin},
  \citenamefont {Ramezani}, \citenamefont {Eichelkraut}, \citenamefont
  {Kottos}, \citenamefont {Cao},\ and\ \citenamefont
  {Christodoulides}}]{UnidirectionalINPTS}%
  \BibitemOpen
  \bibfield  {author} {\bibinfo {author} {\bibfnamefont {Z.}~\bibnamefont
  {Lin}}, \bibinfo {author} {\bibfnamefont {H.}~\bibnamefont {Ramezani}},
  \bibinfo {author} {\bibfnamefont {T.}~\bibnamefont {Eichelkraut}}, \bibinfo
  {author} {\bibfnamefont {T.}~\bibnamefont {Kottos}}, \bibinfo {author}
  {\bibfnamefont {H.}~\bibnamefont {Cao}},\ and\ \bibinfo {author}
  {\bibfnamefont {D.~N.}\ \bibnamefont {Christodoulides}},\ }\bibfield  {title}
  {\bibinfo {title} {{Unidirectional Invisibility Induced by
  $\mathcal{P}\mathcal{T}$-Symmetric Periodic Structures}},\ }\href
  {https://doi.org/10.1103/PhysRevLett.106.213901} {\bibfield  {journal}
  {\bibinfo  {journal} {Phys. Rev. Lett.}\ }\textbf {\bibinfo {volume} {106}},\
  \bibinfo {pages} {213901} (\bibinfo {year} {2011})}\BibitemShut {NoStop}%
\bibitem [{\citenamefont {Feng}\ \emph {et~al.}(2013)\citenamefont {Feng},
  \citenamefont {Xu}, \citenamefont {Fegadolli}, \citenamefont {Lu},
  \citenamefont {Oliveira}, \citenamefont {Almeida}, \citenamefont {Chen},\
  and\ \citenamefont {Scherer}}]{Feng2012_unidirectional}%
  \BibitemOpen
  \bibfield  {author} {\bibinfo {author} {\bibfnamefont {L.}~\bibnamefont
  {Feng}}, \bibinfo {author} {\bibfnamefont {Y.-L.}\ \bibnamefont {Xu}},
  \bibinfo {author} {\bibfnamefont {W.~S.}\ \bibnamefont {Fegadolli}}, \bibinfo
  {author} {\bibfnamefont {M.-H.}\ \bibnamefont {Lu}}, \bibinfo {author}
  {\bibfnamefont {J.~E.~B.}\ \bibnamefont {Oliveira}}, \bibinfo {author}
  {\bibfnamefont {V.~R.}\ \bibnamefont {Almeida}}, \bibinfo {author}
  {\bibfnamefont {Y.-F.}\ \bibnamefont {Chen}},\ and\ \bibinfo {author}
  {\bibfnamefont {A.}~\bibnamefont {Scherer}},\ }\bibfield  {title} {\bibinfo
  {title} {{Experimental demonstration of a unidirectional reflectionless
  parity-time metamaterial at optical frequencies}},\ }\href
  {https://doi.org/10.1038/nmat3495} {\bibfield  {journal} {\bibinfo  {journal}
  {Nature Materials}\ }\textbf {\bibinfo {volume} {12}},\ \bibinfo {pages}
  {108} (\bibinfo {year} {2013})}\BibitemShut {NoStop}%
\bibitem [{\citenamefont {Regensburger}\ \emph {et~al.}(2012)\citenamefont
  {Regensburger}, \citenamefont {Bersch}, \citenamefont {Miri}, \citenamefont
  {Onishchukov}, \citenamefont {Christodoulides},\ and\ \citenamefont
  {Peschel}}]{Regensburger2012_unidirectional}%
  \BibitemOpen
  \bibfield  {author} {\bibinfo {author} {\bibfnamefont {A.}~\bibnamefont
  {Regensburger}}, \bibinfo {author} {\bibfnamefont {C.}~\bibnamefont
  {Bersch}}, \bibinfo {author} {\bibfnamefont {M.~A.}\ \bibnamefont {Miri}},
  \bibinfo {author} {\bibfnamefont {G.}~\bibnamefont {Onishchukov}}, \bibinfo
  {author} {\bibfnamefont {D.~N.}\ \bibnamefont {Christodoulides}},\ and\
  \bibinfo {author} {\bibfnamefont {U.}~\bibnamefont {Peschel}},\ }\bibfield
  {title} {\bibinfo {title} {{Parity-time synthetic photonic lattices}},\
  }\href {https://doi.org/10.1038/nature11298} {\bibfield  {journal} {\bibinfo
  {journal} {Nature}\ }\textbf {\bibinfo {volume} {488}},\ \bibinfo {pages}
  {167} (\bibinfo {year} {2012})}\BibitemShut {NoStop}%
\bibitem [{\citenamefont {Hodaei}\ \emph {et~al.}(2017)\citenamefont {Hodaei},
  \citenamefont {Hassan}, \citenamefont {Wittek}, \citenamefont
  {Garcia-Gracia}, \citenamefont {El-Ganainy}, \citenamefont
  {Christodoulides},\ and\ \citenamefont {Khajavikhan}}]{hodaei2017ESEPPTSB}%
  \BibitemOpen
  \bibfield  {author} {\bibinfo {author} {\bibfnamefont {H.}~\bibnamefont
  {Hodaei}}, \bibinfo {author} {\bibfnamefont {A.~U.}\ \bibnamefont {Hassan}},
  \bibinfo {author} {\bibfnamefont {S.}~\bibnamefont {Wittek}}, \bibinfo
  {author} {\bibfnamefont {H.}~\bibnamefont {Garcia-Gracia}}, \bibinfo {author}
  {\bibfnamefont {R.}~\bibnamefont {El-Ganainy}}, \bibinfo {author}
  {\bibfnamefont {D.~N.}\ \bibnamefont {Christodoulides}},\ and\ \bibinfo
  {author} {\bibfnamefont {M.}~\bibnamefont {Khajavikhan}},\ }\bibfield
  {title} {\bibinfo {title} {{Enhanced sensitivity at higher-order exceptional
  points}},\ }\href {https://doi.org/10.1038/nature23280} {\bibfield  {journal}
  {\bibinfo  {journal} {Nature}\ }\textbf {\bibinfo {volume} {548}},\ \bibinfo
  {pages} {187} (\bibinfo {year} {2017})}\BibitemShut {NoStop}%
\bibitem [{\citenamefont {Chen}\ \emph {et~al.}(2017)\citenamefont {Chen},
  \citenamefont {Kaya~{\"O}zdemir}, \citenamefont {Zhao}, \citenamefont
  {Wiersig},\ and\ \citenamefont {Yang}}]{Sensing2017EPS}%
  \BibitemOpen
  \bibfield  {author} {\bibinfo {author} {\bibfnamefont {W.}~\bibnamefont
  {Chen}}, \bibinfo {author} {\bibfnamefont {{\c{S}}.}~\bibnamefont
  {Kaya~{\"O}zdemir}}, \bibinfo {author} {\bibfnamefont {G.}~\bibnamefont
  {Zhao}}, \bibinfo {author} {\bibfnamefont {J.}~\bibnamefont {Wiersig}},\ and\
  \bibinfo {author} {\bibfnamefont {L.}~\bibnamefont {Yang}},\ }\bibfield
  {title} {\bibinfo {title} {{Exceptional points enhance sensing in an optical
  microcavity}},\ }\href {https://doi.org/10.1038/nature23281} {\bibfield
  {journal} {\bibinfo  {journal} {Nature}\ }\textbf {\bibinfo {volume} {548}},\
  \bibinfo {pages} {192} (\bibinfo {year} {2017})}\BibitemShut {NoStop}%
\bibitem [{\citenamefont {{Zhong, Q. and Ren, J. and Khajavikhan, M. and
  Christodoulides, D. N. and \"Ozdemir, \ifmmode \mbox{\c{S}}\else \c{S}\fi{}.
  K. and El-Ganainy, R.}}(2019)}]{Zhong2019_sensing}%
  \BibitemOpen
  \bibfield  {author} {\bibinfo {author} {\bibnamefont {{Zhong, Q. and Ren, J.
  and Khajavikhan, M. and Christodoulides, D. N. and \"Ozdemir, \ifmmode
  \mbox{\c{S}}\else \c{S}\fi{}. K. and El-Ganainy, R.}}},\ }\bibfield  {title}
  {\bibinfo {title} {{Sensing with Exceptional Surfaces in Order to Combine
  Sensitivity with Robustness}},\ }\href
  {https://doi.org/10.1103/PhysRevLett.122.153902} {\bibfield  {journal}
  {\bibinfo  {journal} {Phys. Rev. Lett.}\ }\textbf {\bibinfo {volume} {122}},\
  \bibinfo {pages} {153902} (\bibinfo {year} {2019})}\BibitemShut {NoStop}%
\bibitem [{\citenamefont {R{\"{u}}ter}\ \emph {et~al.}(2010)\citenamefont
  {R{\"{u}}ter}, \citenamefont {Makris}, \citenamefont {El-Ganainy},
  \citenamefont {Christodoulides}, \citenamefont {Segev},\ and\ \citenamefont
  {Kip}}]{ruter2010PTSOP}%
  \BibitemOpen
  \bibfield  {author} {\bibinfo {author} {\bibfnamefont {C.~E.}\ \bibnamefont
  {R{\"{u}}ter}}, \bibinfo {author} {\bibfnamefont {K.~G.}\ \bibnamefont
  {Makris}}, \bibinfo {author} {\bibfnamefont {R.}~\bibnamefont {El-Ganainy}},
  \bibinfo {author} {\bibfnamefont {D.~N.}\ \bibnamefont {Christodoulides}},
  \bibinfo {author} {\bibfnamefont {M.}~\bibnamefont {Segev}},\ and\ \bibinfo
  {author} {\bibfnamefont {D.}~\bibnamefont {Kip}},\ }\bibfield  {title}
  {\bibinfo {title} {{Observation of parity–-time symmetry in optics}},\
  }\href {https://doi.org/10.1038/nphys1515} {\bibfield  {journal} {\bibinfo
  {journal} {Nature Physics}\ }\textbf {\bibinfo {volume} {6}},\ \bibinfo
  {pages} {192} (\bibinfo {year} {2010})}\BibitemShut {NoStop}%
\bibitem [{\citenamefont {Dragoman}(2011)}]{Dragoman}%
  \BibitemOpen
  \bibfield  {author} {\bibinfo {author} {\bibfnamefont {D.}~\bibnamefont
  {Dragoman}},\ }\bibfield  {title} {\bibinfo {title} {Complex conjugate media:
  Alternative configurations for miniaturized lasers},\ }\href
  {https://doi.org/https://doi.org/10.1016/j.optcom.2010.12.069} {\bibfield
  {journal} {\bibinfo  {journal} {Optics Communications}\ }\textbf {\bibinfo
  {volume} {284}},\ \bibinfo {pages} {2095} (\bibinfo {year}
  {2011})}\BibitemShut {NoStop}%
\bibitem [{\citenamefont {Xu}\ \emph {et~al.}(2017)\citenamefont {Xu},
  \citenamefont {Fu},\ and\ \citenamefont {Chen}}]{Xu:17_CCM}%
  \BibitemOpen
  \bibfield  {author} {\bibinfo {author} {\bibfnamefont {Y.}~\bibnamefont
  {Xu}}, \bibinfo {author} {\bibfnamefont {Y.}~\bibnamefont {Fu}},\ and\
  \bibinfo {author} {\bibfnamefont {H.}~\bibnamefont {Chen}},\ }\bibfield
  {title} {\bibinfo {title} {Electromagnetic wave propagations in conjugate
  metamaterials},\ }\href {https://doi.org/10.1364/OE.25.004952} {\bibfield
  {journal} {\bibinfo  {journal} {Opt. Express}\ }\textbf {\bibinfo {volume}
  {25}},\ \bibinfo {pages} {4952} (\bibinfo {year} {2017})}\BibitemShut
  {NoStop}%
\bibitem [{\citenamefont {Cui}\ \emph {et~al.}(2020)\citenamefont {Cui},
  \citenamefont {Ding}, \citenamefont {Dong},\ and\ \citenamefont
  {Chan}}]{Cui2020:RalizationCCM}%
  \BibitemOpen
  \bibfield  {author} {\bibinfo {author} {\bibfnamefont {X.}~\bibnamefont
  {Cui}}, \bibinfo {author} {\bibfnamefont {K.}~\bibnamefont {Ding}}, \bibinfo
  {author} {\bibfnamefont {J.-W.}\ \bibnamefont {Dong}},\ and\ \bibinfo
  {author} {\bibfnamefont {C.}~\bibnamefont {Chan}},\ }\bibfield  {title}
  {\bibinfo {title} {{Realization of complex conjugate media using
  non-PT-symmetric photonic crystals}},\ }\href
  {https://doi.org/doi:10.1515/nanoph-2019-0389} {\bibfield  {journal}
  {\bibinfo  {journal} {Nanophotonics}\ }\textbf {\bibinfo {volume} {9}},\
  \bibinfo {pages} {195} (\bibinfo {year} {2020})}\BibitemShut {NoStop}%
\bibitem [{\citenamefont {Basiri}\ \emph {et~al.}(2015)\citenamefont {Basiri},
  \citenamefont {Vitebskiy},\ and\ \citenamefont {Kottos}}]{Basiri2015:CCM}%
  \BibitemOpen
  \bibfield  {author} {\bibinfo {author} {\bibfnamefont {A.}~\bibnamefont
  {Basiri}}, \bibinfo {author} {\bibfnamefont {I.}~\bibnamefont {Vitebskiy}},\
  and\ \bibinfo {author} {\bibfnamefont {T.}~\bibnamefont {Kottos}},\
  }\bibfield  {title} {\bibinfo {title} {{Light scattering in pseudopassive
  media with uniformly balanced gain and loss}},\ }\href
  {https://doi.org/10.1103/PhysRevA.91.063843} {\bibfield  {journal} {\bibinfo
  {journal} {Phys. Rev. A}\ }\textbf {\bibinfo {volume} {91}},\ \bibinfo
  {pages} {063843} (\bibinfo {year} {2015})}\BibitemShut {NoStop}%
\bibitem [{\citenamefont {Fu}\ \emph {et~al.}(2017)\citenamefont {Fu},
  \citenamefont {Cao}, \citenamefont {Cummer}, \citenamefont {Xu},\ and\
  \citenamefont {Chen}}]{PhysRevA.96.043838_CPA_laser}%
  \BibitemOpen
  \bibfield  {author} {\bibinfo {author} {\bibfnamefont {Y.}~\bibnamefont
  {Fu}}, \bibinfo {author} {\bibfnamefont {Y.}~\bibnamefont {Cao}}, \bibinfo
  {author} {\bibfnamefont {S.~A.}\ \bibnamefont {Cummer}}, \bibinfo {author}
  {\bibfnamefont {Y.}~\bibnamefont {Xu}},\ and\ \bibinfo {author}
  {\bibfnamefont {H.}~\bibnamefont {Chen}},\ }\bibfield  {title} {\bibinfo
  {title} {Coherent perfect absorber and laser modes in purely imaginary
  metamaterials},\ }\href {https://doi.org/10.1103/PhysRevA.96.043838}
  {\bibfield  {journal} {\bibinfo  {journal} {Phys. Rev. A}\ }\textbf {\bibinfo
  {volume} {96}},\ \bibinfo {pages} {043838} (\bibinfo {year}
  {2017})}\BibitemShut {NoStop}%
\bibitem [{\citenamefont {Bai}\ \emph {et~al.}(2016)\citenamefont {Bai},
  \citenamefont {Ding}, \citenamefont {Wang}, \citenamefont {Luo},
  \citenamefont {Zhang}, \citenamefont {Chan}, \citenamefont {Wu},\ and\
  \citenamefont {Lai}}]{Bai2016:CCM_absorber}%
  \BibitemOpen
  \bibfield  {author} {\bibinfo {author} {\bibfnamefont {P.}~\bibnamefont
  {Bai}}, \bibinfo {author} {\bibfnamefont {K.}~\bibnamefont {Ding}}, \bibinfo
  {author} {\bibfnamefont {G.}~\bibnamefont {Wang}}, \bibinfo {author}
  {\bibfnamefont {J.}~\bibnamefont {Luo}}, \bibinfo {author} {\bibfnamefont
  {Z.-Q.}\ \bibnamefont {Zhang}}, \bibinfo {author} {\bibfnamefont {C.~T.}\
  \bibnamefont {Chan}}, \bibinfo {author} {\bibfnamefont {Y.}~\bibnamefont
  {Wu}},\ and\ \bibinfo {author} {\bibfnamefont {Y.}~\bibnamefont {Lai}},\
  }\bibfield  {title} {\bibinfo {title} {{Simultaneous realization of a
  coherent perfect absorber and laser by zero-index media with both gain and
  loss}},\ }\href {https://doi.org/10.1103/PhysRevA.94.063841} {\bibfield
  {journal} {\bibinfo  {journal} {Phys. Rev. A}\ }\textbf {\bibinfo {volume}
  {94}},\ \bibinfo {pages} {063841} (\bibinfo {year} {2016})}\BibitemShut
  {NoStop}%
\bibitem [{\citenamefont {Xu}\ \emph {et~al.}(2021)\citenamefont {Xu},
  \citenamefont {Farhat},\ and\ \citenamefont {Wu}}]{Xu2021:CCM}%
  \BibitemOpen
  \bibfield  {author} {\bibinfo {author} {\bibfnamefont {C.}~\bibnamefont
  {Xu}}, \bibinfo {author} {\bibfnamefont {M.}~\bibnamefont {Farhat}},\ and\
  \bibinfo {author} {\bibfnamefont {Y.}~\bibnamefont {Wu}},\ }\bibfield
  {title} {\bibinfo {title} {Non-hermitian electromagnetic double-near-zero
  index medium in a two-dimensional photonic crystal},\ }\href
  {https://doi.org/10.1063/5.0073391} {\bibfield  {journal} {\bibinfo
  {journal} {Applied Physics Letters}\ }\textbf {\bibinfo {volume} {119}},\
  \bibinfo {pages} {224102} (\bibinfo {year} {2021})}\BibitemShut {NoStop}%
\bibitem [{\citenamefont {Mirmoosa}\ \emph {et~al.}(2023)\citenamefont
  {Mirmoosa}, \citenamefont {Wang}, \citenamefont {Freter}, \citenamefont
  {Sihvola},\ and\ \citenamefont {Tretyakov}}]{mirmoosaTellegen}%
  \BibitemOpen
  \bibfield  {author} {\bibinfo {author} {\bibfnamefont {M.~S.}\ \bibnamefont
  {Mirmoosa}}, \bibinfo {author} {\bibfnamefont {X.}~\bibnamefont {Wang}},
  \bibinfo {author} {\bibfnamefont {L.}~\bibnamefont {Freter}}, \bibinfo
  {author} {\bibfnamefont {A.}~\bibnamefont {Sihvola}},\ and\ \bibinfo {author}
  {\bibfnamefont {S.}~\bibnamefont {Tretyakov}},\ }\bibfield  {title} {\bibinfo
  {title} {{Loss\textendash gain compensated anti-Hermitian magnetodielectric
  medium to realize Tellegen nihility effects}},\ }\href
  {https://doi.org/10.1364/OL.483103} {\bibfield  {journal} {\bibinfo
  {journal} {Opt. Lett.}\ }\textbf {\bibinfo {volume} {48}},\ \bibinfo {pages}
  {1032} (\bibinfo {year} {2023})}\BibitemShut {NoStop}%
\bibitem [{\citenamefont {Glybovski}\ \emph {et~al.}(2016)\citenamefont
  {Glybovski}, \citenamefont {Tretyakov}, \citenamefont {Belov}, \citenamefont
  {Kivshar},\ and\ \citenamefont {Simovski}}]{Glybovski}%
  \BibitemOpen
  \bibfield  {author} {\bibinfo {author} {\bibfnamefont {S.~B.}\ \bibnamefont
  {Glybovski}}, \bibinfo {author} {\bibfnamefont {S.~A.}\ \bibnamefont
  {Tretyakov}}, \bibinfo {author} {\bibfnamefont {P.~A.}\ \bibnamefont
  {Belov}}, \bibinfo {author} {\bibfnamefont {Y.~S.}\ \bibnamefont {Kivshar}},\
  and\ \bibinfo {author} {\bibfnamefont {C.~R.}\ \bibnamefont {Simovski}},\
  }\bibfield  {title} {\bibinfo {title} {{Metasurfaces: From microwaves to
  visible}},\ }\href
  {https://doi.org/https://doi.org/10.1016/j.physrep.2016.04.004} {\bibfield
  {journal} {\bibinfo  {journal} {Physics Reports}\ }\textbf {\bibinfo {volume}
  {634}},\ \bibinfo {pages} {1} (\bibinfo {year} {2016})}\BibitemShut {NoStop}%
\bibitem [{\citenamefont {Bohren}\ and\ \citenamefont
  {Huffman}(1983)}]{Bohren-Huffman}%
  \BibitemOpen
  \bibfield  {author} {\bibinfo {author} {\bibfnamefont {C.~F.}\ \bibnamefont
  {Bohren}}\ and\ \bibinfo {author} {\bibfnamefont {D.~R.}\ \bibnamefont
  {Huffman}},\ }\href@noop {} {\emph {\bibinfo {title} {Absorption and
  Scattering of Light by Small Particles}}}\ (\bibinfo  {publisher} {Wiley},\
  \bibinfo {address} {New York},\ \bibinfo {year} {1983})\BibitemShut {NoStop}%
\bibitem [{\citenamefont {Tzarouchis}\ and\ \citenamefont
  {Sihvola}(2018)}]{Ari_Mie}%
  \BibitemOpen
  \bibfield  {author} {\bibinfo {author} {\bibfnamefont {D.}~\bibnamefont
  {Tzarouchis}}\ and\ \bibinfo {author} {\bibfnamefont {A.}~\bibnamefont
  {Sihvola}},\ }\bibfield  {title} {\bibinfo {title} {{Light Scattering by a
  Dielectric Sphere: Perspectives on the Mie Resonances}},\ }\bibfield
  {journal} {\bibinfo  {journal} {Applied Sciences}\ }\textbf {\bibinfo
  {volume} {8}},\ \href {https://doi.org/10.3390/app8020184}
  {10.3390/app8020184} (\bibinfo {year} {2018})\BibitemShut {NoStop}%
\bibitem [{\citenamefont {Osipov}\ and\ \citenamefont
  {Tretyakov}(2017)}]{Osipov_book}%
  \BibitemOpen
  \bibfield  {author} {\bibinfo {author} {\bibfnamefont {A.}~\bibnamefont
  {Osipov}}\ and\ \bibinfo {author} {\bibfnamefont {S.}~\bibnamefont
  {Tretyakov}},\ }in\ \href@noop {} {\emph {\bibinfo {booktitle} {Modern
  Electromagnetic Scattering Theory with Applications}}}\ (\bibinfo
  {publisher} {John Wiley \& Sons, UK},\ \bibinfo {year} {2017})\BibitemShut
  {NoStop}%
\bibitem [{\citenamefont {Kerker}\ \emph {et~al.}(1983)\citenamefont {Kerker},
  \citenamefont {Wang},\ and\ \citenamefont {Giles}}]{Kerker83}%
  \BibitemOpen
  \bibfield  {author} {\bibinfo {author} {\bibfnamefont {M.}~\bibnamefont
  {Kerker}}, \bibinfo {author} {\bibfnamefont {D.-S.}\ \bibnamefont {Wang}},\
  and\ \bibinfo {author} {\bibfnamefont {C.~L.}\ \bibnamefont {Giles}},\
  }\bibfield  {title} {\bibinfo {title} {Electromagnetic scattering by magnetic
  spheres},\ }\href {https://doi.org/10.1364/JOSA.73.000765} {\bibfield
  {journal} {\bibinfo  {journal} {J. Opt. Soc. Am.}\ }\textbf {\bibinfo
  {volume} {73}},\ \bibinfo {pages} {765} (\bibinfo {year} {1983})}\BibitemShut
  {NoStop}%
\bibitem [{\citenamefont {Alu}\ and\ \citenamefont
  {Engheta}(2010)}]{Alu_Engheta_2010}%
  \BibitemOpen
  \bibfield  {author} {\bibinfo {author} {\bibfnamefont {A.}~\bibnamefont
  {Alu}}\ and\ \bibinfo {author} {\bibfnamefont {N.}~\bibnamefont {Engheta}},\
  }\bibfield  {title} {\bibinfo {title} {{How does zero forward-scattering in
  magnetodielectric nanoparticles comply with the optical theorem?}},\ }\href
  {https://doi.org/10.1117/1.3449103} {\bibfield  {journal} {\bibinfo
  {journal} {Journal of Nanophotonics}\ }\textbf {\bibinfo {volume} {4}},\
  \bibinfo {pages} {041590} (\bibinfo {year} {2010})}\BibitemShut {NoStop}%
\bibitem [{\citenamefont {Sihvola}\ \emph {et~al.}(2021)\citenamefont
  {Sihvola}, \citenamefont {Parveen}, \citenamefont {Wall{\'e}n},\ and\
  \citenamefont {Yl{\"a}-Oijala}}]{ARISihvola2021EBAP}%
  \BibitemOpen
  \bibfield  {author} {\bibinfo {author} {\bibfnamefont {A.}~\bibnamefont
  {Sihvola}}, \bibinfo {author} {\bibfnamefont {R.}~\bibnamefont {Parveen}},
  \bibinfo {author} {\bibfnamefont {H.}~\bibnamefont {Wall{\'e}n}},\ and\
  \bibinfo {author} {\bibfnamefont {P.}~\bibnamefont {Yl{\"a}-Oijala}},\
  }\bibfield  {title} {\bibinfo {title} {{Enhanced backscattering for
  dielectrically active scatterers}},\ }\href
  {https://doi.org/10.46620/21-0020} {\bibfield  {journal} {\bibinfo  {journal}
  {URSI Radio Sci. Lett.}\ }\textbf {\bibinfo {volume} {3}},\ \bibinfo {pages}
  {1} (\bibinfo {year} {2021})}\BibitemShut {NoStop}%
\end{thebibliography}%



\end{document}